# Intralayer Carbon Substitution in the $MgB_2$ Superconductor

T. Takenobu,[1] T. Ito,[1] Dam H. Chi,[1] K. Prassides,[1,2,3] and Y. Iwasa[1]

[1] *Japan Advanced Institute of Science and Technology, Tatsunokuchi, Ishikawa 923-1292, Japan*
[2] *School of Chemistry, Physics and Environmental Science, University of Sussex, Brighton BN1 9QJ, UK*
[3] *Institute of Materials Science, NCSR "Demokritos", 15310 Ag. Paraskevi, Athens, Greece*

**Abstract**

We report that the ternary $MgB_{2-x}C_x$ compounds adopt an isostructural $AlB_2$-type hexagonal structure in a relatively small range of nominal carbon concentration, $x< 0.1$. The lattice parameter *a* decreases almost linearly with increasing carbon content *x*, while the *c* parameter remains unchanged, indicating that carbon is exclusively substituted in the boron honeycomb layer without affecting the interlayer interactions. The superconducting transition temperature $T_c$, determined by magnetometry experiments, also decreases quasilinearly as a function of the carbon concentration. The structural and electronic behavior of $MgB_{2-x}C_x$ displays a remarkable similarity with the isoelectronic $Mg_{1-x}Al_xB_2$ despite the different substitution sites.

Discovery of superconductivity in $MgB_2$ at $T_c$= 39 K (1) is attracting wide attention because of the simplicity in the chemical composition, the crystal and electronic structure of the system and its highly promising potential applications. Detailed information on the properties of $MgB_2$, particularly related to the nature of superconductivity, is being currently rapidly accumulated by means of structural and electronic probes on the parent compound, $MgB_2$. An alternative approach is to synthesize related materials by partial chemical substitution on either the Mg or the B interleaved layers and follow the evolution of the properties.

$MgB_2$ adopts a hexagonal structure ($AlB_2$-type, space group *P*6/*mmm*) (2) which is analogous to intercalated graphite with all hexagonal prismatic sites of the primitive graphitic structure (found in hexagonal BN) completely filled and resulting in two interleaved B and Mg layers. In addition, allowing for full charge transfer from Mg to the boron 2D sheets, the latter are themselves isoelectronic with graphite. Also, it has been known that various metal borides form an isostructural series of compounds (*2*). Moreover, theoretical calculations predict that substitution of Mg results in significant changes of the density of states at the Fermi level without introducing any disorder, potentially allowing access to increased $T_c$ (3-5). These structural features have motivated attempts to substitute Mg with alkali (6), alkaline-earth, group III metals (7,8), and other elements. However, a significant difference of $MgB_2$ from graphite intercalation compounds is that both the structural and electronic properties are substantially more isotropic. For example, high pressure synchrotron X-ray diffraction experiments revealed that the isothermal compressibility of $MgB_2$ is only moderately anisotropic between the boron intra- and inter-layer directions (9,10). Also, band structure calculations showed that the electronic states of this superconductor are essentially three-dimensional (3,4,5,11). Thus, the substitution of Mg sites is not entirely analogous to the case of intercalation of the strongly bonded strictly 2D graphitic sheets (12). Instead, appropriate substitution on the more weakly bonded boron sheets offers an alternative route of modulating the structural and electronic properties of this system. In this letter, we report the synthesis of carbon-doped $MgB_2$ superconductors, $MgB_{2-x}C_x$ and their solid solution behavior in the range of carbon concentrations $0 \leq x \leq 0.1$. The observed structural and superconducting properties of $MgB_{2-x}C_x$ ternaries display an amazing coincidence to the isoelectronic $Mg_{1-x}Al_xB_2$ series, despite the difference in substitution sites. These results indicate that the electronic structure of $MgB_2$ and related ternary systems is well described by the band theory and that $T_c$ is controlled by the density of states at the Fermi level.

The $MgB_{2-x}C_x$ ($x$= 0.02, 0.04, 0.06, 0.1, 0.2, 0.5) samples were synthesized by heating mixed powders of amorphous boron, carbon black, and magnesium at 900$^o$C for 2 hours. The powders are placed in stainless steel tubes and sealed inside quartz tubes. Figure 1 shows the (002) and ($\bar{1}10$) Bragg reflections taken on an X-ray powder diffractometer with Cu-$K_\alpha$ radiation for $x$= 0.0, 0.02, 0.04, 0.06, and 0.1. While the position of the (002) peak remains unchanged with increasing $x$, the ($\bar{1}10$) peak shifts continuously to higher angles up to $x$= 0.06. Other reflection peaks with different combinations of Miller indices show consistent behavior with these typical reflections. At $x$= 0.1, the diffraction peaks display a considerable broadening, which is attributable to the reduction in crystallinity or to the onset of phase separation. The diffraction profiles of the nominal $x$= 0.2 and 0.5 compositions clearly show a two-phase behavior with a large number of extra peaks which cannot be accounted for within the $AlB_2$-type structure.

The lattice parameters were extracted by a Le Bail pattern decomposition technique (13). Figure 2 shows the evolution of the hexagonal lattice constants with carbon concentration. The $a$ and $c$ lattice parameters display strongly contrasting behavior: $a$ contracts essentially in a linear fashion up to $x$= 0.06, while $c$ is invariant with $x$, unambiguously indicating that carbon is substituted in the boron layers without affacting their interlayer separation. This type of substitution effect has been also encountered in Li doped $MgB_2$ (6), but is in sharp contrast with the results of Al substitution (7), in which the in-plane B-B separation is essentially invariant. The homogeneous and random nature of the carbon substitution is also evident in the linewidths of the reflections which remain sharp and are independent of $x$ up to $x$= 0.06. On the other hand, an abrupt broadening occurs at $x$= 0.1 implying incipient sample inhomogeneity. Such inhomogenity could be the signature of the onset of phase separation, as multiphase behavior is encountered for the $x$= 0.2 and 0.5 samples. The structual data provide unambiguous evidence of solid solution behavior, with an extremely small miscibility region of 0< $x$< 0.1.

Magnetic susceptibility of the compositions with $x$= 0.0, 0.02, 0.04, 0.06, 0.1, 0.2, and 0.5 has been measured with a Quantum Design SQUID magnetometer. Figure 3 displays the temperature dependence of the susceptibility in zero-field cooled experiments at 10 Oe. All the samples up to $x$= 0.1 show well defined one-step transitions and shielding fractions of above 100% at 10 K before correction for demagnetization effect, implying that superconductivity is of bulk nature. The transition temperature, $T_c$, defined by the intersection of line extrapolations made both below and above $T_c$, decreases with increasing $x$ at the rate of -d$T_c$/d$x$= 57 K. Above $x$= 0.1, the transition becomes too broad to allow accurate definition of $T_c$ and the volume fraction continuously decreases. The increase of $T_c$, suggested by the resistivity measurements on the multiphase $x$= 0.2 composition (12) is not observed in the present bulk characterization.

Fig. 4 displays the variation of $T_c$ with $x$ in the solid solution region for $MgB_{2-x}C_x$. The results for $Mg_{1-x}Al_xB_2$ (7) are also included (dotted line) for comparison. We note that as C has one more electron than B, electron doping is also anticipated in $MgB_{2-x}C_x$ in direct analogy with the $Mg_{1-x}Al_xB_2$ case. However, the differing substitution site and resulting lattice response allow us to compare the effects of the two distinct layers (boron *vs* magnesium) on the properties of the system. Band structure calculations predict a reduction in the density of states (DOS) at the Fermi level for both $Mg_{1-x}Al_xB_2$ and $MgB_{2-x}C_x$ (14,15). The observed decrease in $T_c$ for both systems can then be rationalized as a direct result of this reduction. However, we note from Figure 4 that the rate of decrease in $T_c$ is considerably larger in $MgB_{2-x}C_x$ (2-3 times that in $Mg_{1-x}Al_xB_2$). One possibility is that this reflects the details of the proposed somewhat anisotropic electronic structure of $MgB_2$, in which electronic conduction is mostly dominated by the boron sheets and could be more sensitive to disorder effects in, rather than between, the layers. However, a relevant comparison between the two systems should involve the lattice constant coefficients of $T_c$, taking into account that in $MgB_{2-x}C_x$, the $c$ axis remains unchanged, while in $Mg_{1-x}Al_xB_2$, the boron intraplane distances do not change. It is then intriguing to note that the extracted coefficients, -dln$T_c$/d$a$ ($MgB_{2-x}C_x$) and –dln$T_c$/d$c$ ($Mg_{1-x}Al_xB_2$) are essentially identical, on the order of 7 $Å^{-1}$. This implies a very similar dependence irrespective of whether substitution occurs in or between the boron sheets, consistent with an almost isotropic nature of the conduction band.

The results presented in this paper indicate that the electronic structure of $MgB_2$ and related ternary systems is well described by the band theory and that the value of $T_c$ is controlled by the density of states at the Fermi level, consistent with a conventional BCS-type origin of superconductivity. Despite the reduced $T_c$, the availability of families of well-characterized electron-doped $MgB_2$ compositions should allow the systematic study of the electronic properties of these intriguing superconductors with the variation of the doping level.

**Acknowledgements**. KP thanks Monbusho for a Visiting Professorship to JAIST. We thank C. J. Nuttall for valuable discussion. We acknowledge H. Iwasaki and T. Naito for the provision of SQUID machine time.

## References and Notes

1. J. Nagamatsu, N. Nakagawa, T. Muranaka, Y. Zenitani, and J. Akimitsu, *Nature*, 410, 63 (2001).
2. M. E. Jones and R. E. Marsh, *J. Am. Chem. Soc*. **76**, 1434 (1954).
3. J. E. Hirsch, cond-mat/0102115.
4. S. Suzuki, S. Higai, and K. Nakao, *J. Phys. Soc. Jpn*., submitted
5. G. Satta, G. Profeta, F. Bernardini, A. Continenza, and S. Massidda, cond-mat/0102358.
6. Y. G. Zao, X. P. Zhang, P. T. Qiao, H. T. Zhang, S. L. Jia, B. S. Cao, M. H. Zhu, Z. H. Han, X. L. Wang, and B. L. Gu, cond-mat/0103007 (2001).
7. J. S. Slusky, N. Rogado, K. A. Regan, M. A. Heyward, P. Khalifa, T. He, K. Inumaru, S. Loureiro, M. K. Haas, H. Zandbergen, and R. J. Cava, cond-mat/0102262 (2001).
8. I. Felner, cond-mat/0102508 (2001).
9. K. Prassides, Y. Iwasa, T. Ito, Dam H. Chi, K. Uehara, E. Nishibori, M. Takata, M. Sakata, Y. Ohishi, O. Shimomura, T. Muranaka, and J. Akimitsu, cond-mat/0102507.
10. T. Vogt, G. Schneider, J. A. Hriljac, G. Yang, and J. S. Abel, cond-mat/0102480.
11. I. Loa and K. Syassen, Solid State Commun., in press.
12. J. S. Ahn and E. J. Choi, cond-mat/0103169 (2001).
13. J. Rodriguez-Carvajal, Program Fullprof (Version 3.5d-Oct. 98), LLB-JRC (unpublished).
14. A. L. Ivanovskii and N. I. Medvedeva, *Russian J. Inorg. Chem*. **45**, 1234 (2000).
15. N. I. Medvedeva, A. L. Ivanovskii, J. E. Medvedeva, and A. J. Freeman, cond-mat/0103157 (2001).

**Figure captions**

**Fig. 1** The ($\bar{1}10$) and (002) Bragg reflections for the $MgB_{2-x}C_x$ compositions ($x$= 0.0, 0.02, 0.04, 0.06, and 0.1) .

**Fig. 2** Lattice parameters $a$ and $c$ as a function of the nominal carbon content, $x$. The shaded area indicates a region showing inhomogenity of samples.

**Fig. 3** Zero-field cooled magnetic susceptibility measured at 10 Oe for the $MgB_{2-x}C_x$ compounds ($x$= 0.0, 0.02, 0.04, 0.06, and 0.1). The arrows indicate $T_c$'s for x=0.0, 0.02, 0.04, and 0.06.

**Fig. 4** $T_c$ vs. substitution concentration that corresponds to the excess electron count for $MgB_{2-x}C_x$ (filled circle) and $Mg_{1-x}Al_xB_2$ (dotted line) taken from Ref. 7.

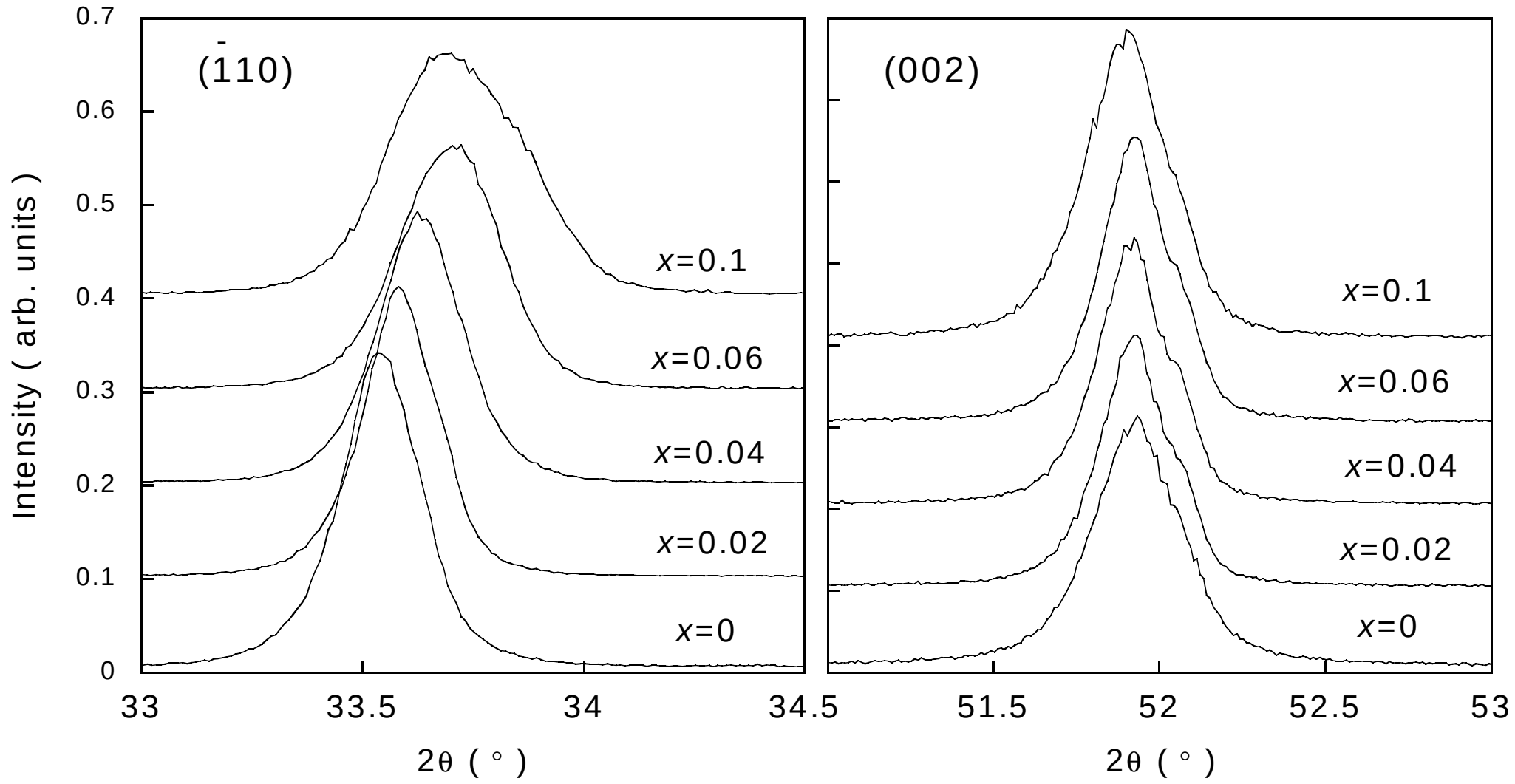

Fig. 1 Takenobu *et al*.

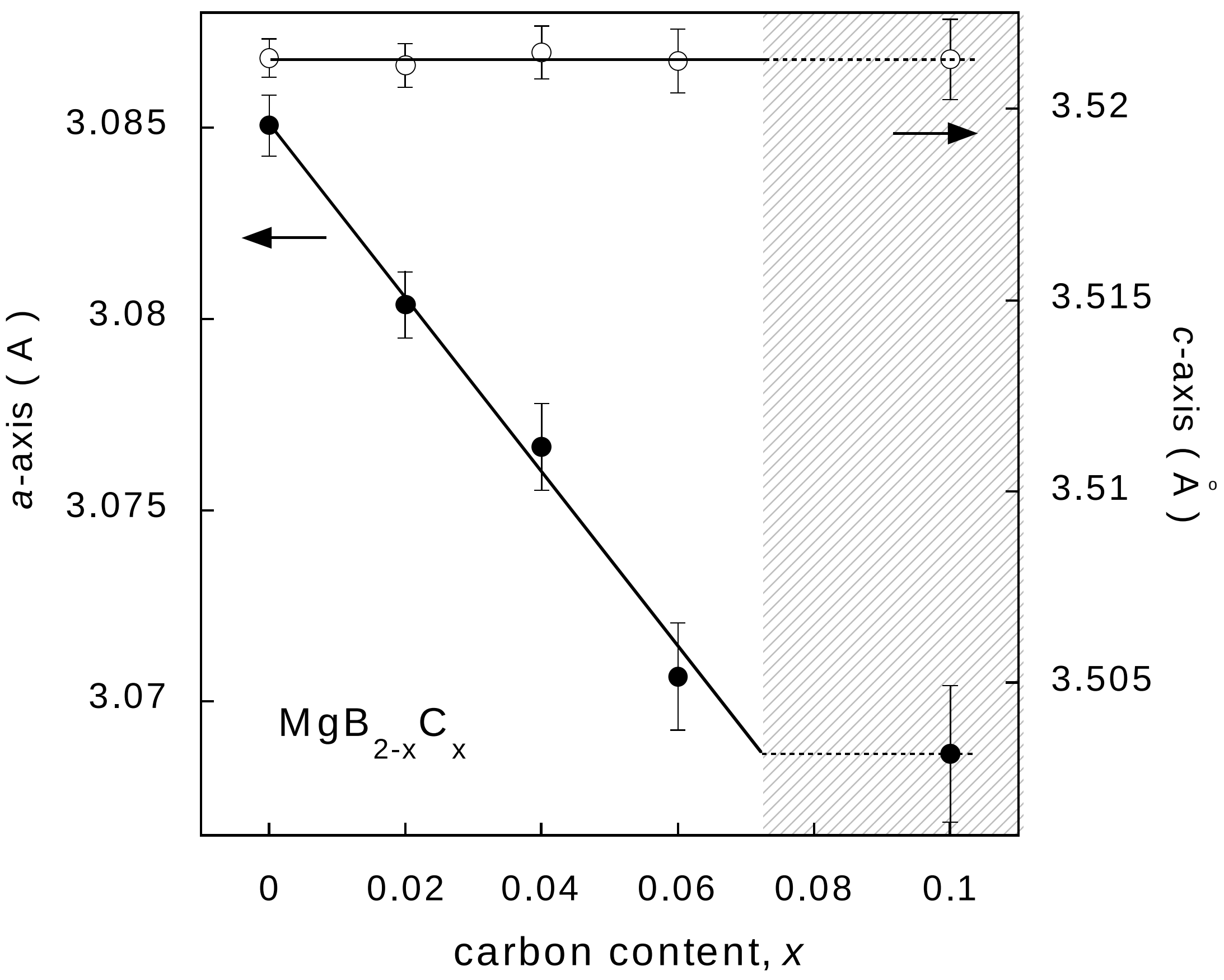

Fig. 2 Takenobu *et al*.

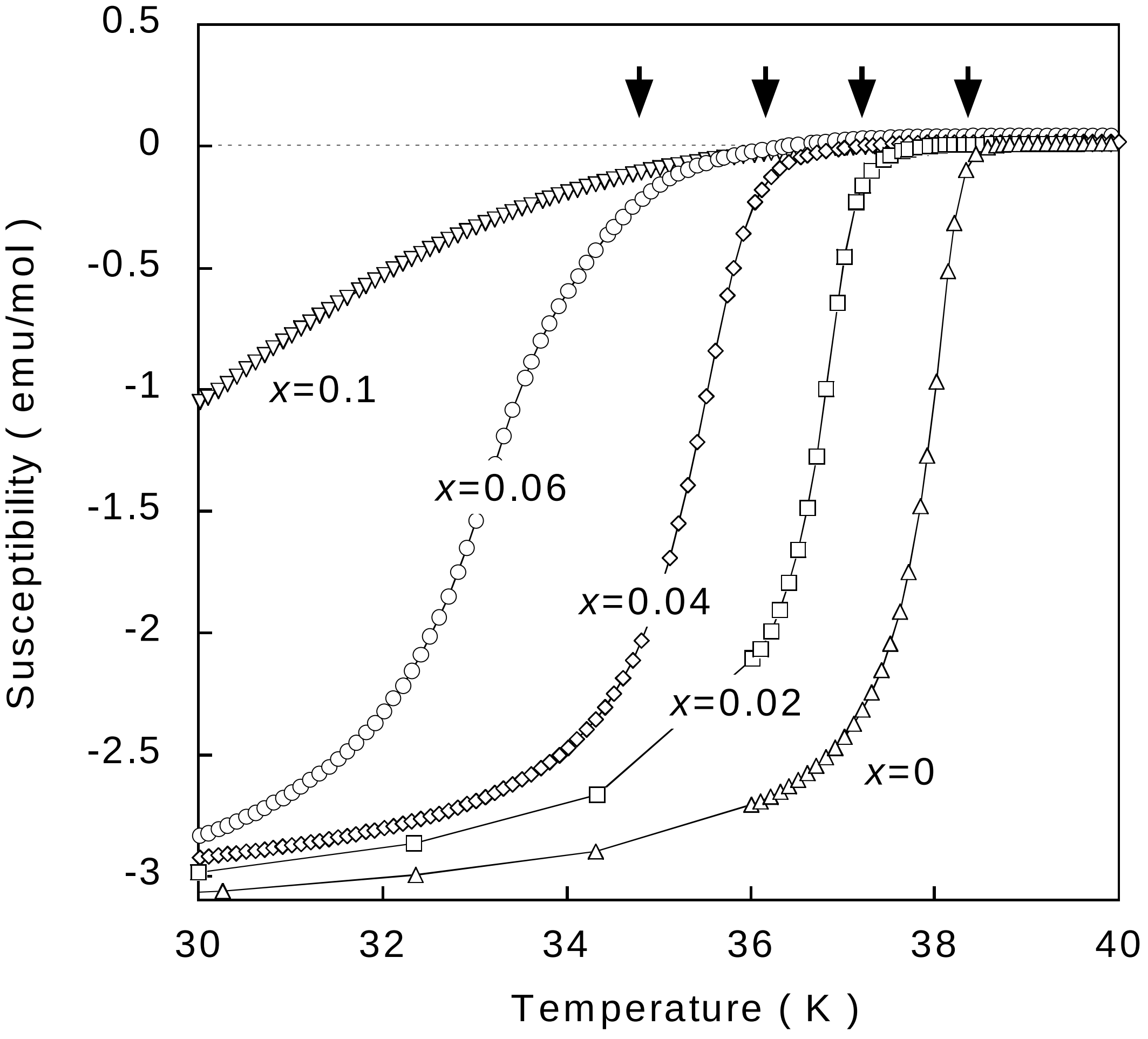

Fig. 3 Takenobu *et al*.

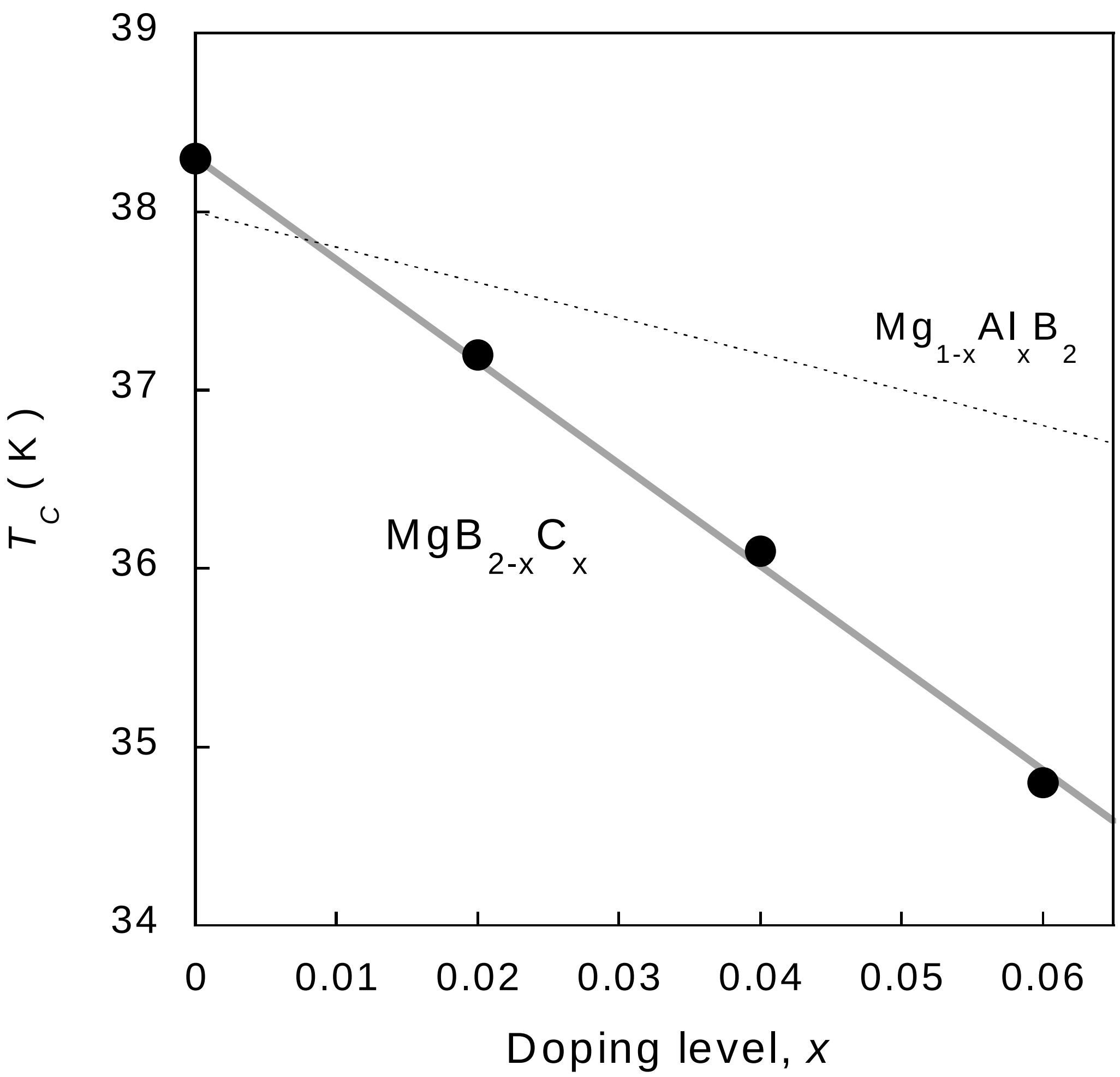

Fig. 4 Takenobu *et al*.